**Quantitatively analyzing the mechanism of giant circular dichroism in extrinsic plasmonic chiral nanostructures by the interplay of electric and magnetic dipoles**

Li Hu[1,4,†], Xiaorui Tian[2,†], Yingzhou Huang[1], Xinqiang Wang[1] and Yurui Fang[3,*]

[1]Soft Matter and Interdisciplinary Research Center, College of Physics, Chongqing University, Chongqing, 400044, P. R. China

[2]College of Chemistry, Chemical Engineering and Materials Science, Shandong Normal University, Jinan 250014, China

[3]Bionanophotonics, Department of Applied Physics, Chalmers University of Technology, Göteborg, SE-41296, Sweden.

[4]School of Computer Science and Information Engineering, Chongqing Technology and Business University, Chongqing, 400067, China

†These authors contribute equally

*Corresponding Email: yurui.fang@chalmers.se (Y. Fang)

**Abstract**

The plasmonic chirality has drawn a lot of attention because of the tunable circular dichroism (CD) and the enhancement for the signal of chiral molecules. Different mechanisms have been proposed for explaining the plasmonic CD, however, a quantitative one like ab initio mechanism in chiral molecules is still unavailable. In this work, a mechanism similar to the chiral molecules is analyzed. The giant extrinsic circular dichroism of plasmonic splitting rectangle ring is quantitatively investigated theoretically. The interplay of electric and magnetic modes of the meta-structure is proposed to explain the giant CD. The interplay is analyzed both in an analytical coupled electric-magnetic dipoles model and finite element method model. The surface charge distributions show that the circular current yielded in the splitting rectangle ring makes it behave like a magneton at some resonant modes, which interact with electric modes and results in a mixing of the two kinds of modes. The strong interplay of the two kinds of modes is mainly responsible for the giant CD.The analysis of the chiral near field of the structure shows potential applications in chiral molecule sensing.

## Introduction

The optical chirality due to different responses of the handedness or asymmetry of the object to the spin of light is very weak but very important in nature.[1] The analysis of chiral effects is of significance in the study of medicine diagnose, crystallography, analytical chemistry, molecular biology and life forms in universe.[2] However, the weak chiral response of nature molecules has limited those applications, especially when the probed target is tiny. One recent way is designing metal nanostructures to enhance or yield such effect, where surface plasmons (SPs), the collective oscillation of free electrons, play key roles. The surface plasmon resonance is sensitively dependent on the shape, size, material, as well as the configuration of nanostructures, which make it be flexibly manipulated and have applications in various areas[3]. Metal nanostructures have been considered as good candidate for associating the chiral effect and extenting existence of chirality in our nature.[4-6]

As an emerging area, optical activity of plasmonic nanostructures has attracted increasing attentions in recent years. Optical activities such as optical rotatory dispersion (ORD) and circular dichroism (CD) are produced by the mirror asymmetry (chirality) of structures. A pure two dimensional asymmetric structure under normal incidence excitation is thought only having birefringence effects, which only yield ORD. While chiral structures that can yield CD should be three dimensional, at leastin some way (e.g. substrate image effect is also ok),[7] so as that their mirror forms cannot be superimposed with themselves through a sequence of rotations and translations.[1] In the past several years, various chiral plasmonic nanostructures have been investigated for their optical activities, such as chiral metal particles,[8, 9] pairs of mutually twisted planar metal patterns,[10] single-layered metal sawtooth gratings,[11] planar chiral metal patterns,[12] DNA based self-assembled metal particles,[13, 14] metal helix,[15-17] etc. As a straightforward geometrical definition, chirality can be obtained from achiral nanostructures simply by tilting theirsymmetry axis out of the incident plane to yield the three dimensional asymmetry, which is so-called extrinsic chirality.[18-20] Giant chirality has been obtained with such extrinsic chiral structures since 2009.[18] The extrinsic chirality not only provides more flexible way to overcome the difficulty in fabrication progress of complex chiral structures, but also shows even stronger CD than the intrinsic one. A near 100% difference of response to the spin light was theoretically predicted last year in extrinsic chirality of plasmonic nanorice heterodimers due to the Fano associated effect.[21]

However, by now almost all of the explanations for the CD effect of plasmonic structures are still phenomenological. Someone explained as the phase delay of two orthogonal directions of the structures, someone explained as the asymmetry of the structure to the circularly polarized light (CPL) and someone explained as the resonant modes overlapping with the rotation direction of the CPL. Mostly, the phase changing is considered qualitatively. A more general profound way is analogic to the plasmonic structure as a chiral molecule, whose chiral mechanism has been well investigated and explained with the ab initio theory. In the chiral molecule model, the chiral mechanism is obtained from ab initio model that the chiral effect comes from the interaction of the electric and magnetic dipoles.[1] The idea was borrowed by Plum,[18, 20] and qualitatively explained the extrinsic chirality of plasmonic structures. Later, Tang[22] borrowed Lipkin's theory[23] and developed the chirality of the chiral field. The chirality of plasmonic structures and chirality enhancement of plasmonic structures can be well explained with combining these two ideas[21, 24-26]. However, the analytical or quantitative explanation of the CD is still unavailable.

In this work, we present an analytical model and quantitatively analyze the giant CD of plasmonic splitting rectangle rings theoretically. The interplay of electric and magnetic modes of the meta-structure is analyzed and considered to be responsible for the giant CD. The results show that the circular current yielded in the splitting rectangle ring behaves like a magneton. The hybridization of the electric mode and magnetic mode results in a mixing of the two modes. The analytical model as well as the finite element method (FEM, COMSOL multiphysics) analysis matches the CD quite well quantitatively. In addition, splitting rectangle rings of different parameters are further investigated. The analysis of the chiral near fields of the structure shows potential applications in chiral molecule sensing.

## Model

The structure investigated in this work is a well-known splitting rectangle ring shown in Figure 1, which has been investigated a lot in the past years.[27-29] We consider the gold splitting rectangle ring (optical constants from Johnson and Christy) located in the x-y plane and the incident light propagating in the y-z plane with $\theta$ representing the angle of **k** off the z axis and φ for the angle off the x axis. The length and width of the rectangular ring were set to the same value of $l$, the thickness and the right asymmetric arm set as $h$ and $a$, the gap set as $d$, each of which is marked in Figure 1a&b. All calculations were performed assuming the structure in a homogeneous surrounding medium with effective refractive index of 1.1. The circular dichroism of the system were calculated as the difference in extinction under left- and right-handed circularly polarized light ($CD = \sigma_L - \sigma_R$ ). To satisfy the extrinsically chiral excitation criteria, the splitting gap cannot be in the middle of the long side, here setting $a$ = 20 nm. Oblique excitation with wavevector **k** not in the plane determined by the normal and symmetric axis of the structure will cause CD effect. Here it can be seen that when excited under the case of Figure 1b, huge CD can be obtained in the dipole peak, which shows more than 90% difference (Figure 1c and 1d). Such huge cd in dipole mode is seldom reported. As indicated, there is no optical activity under normal incidence ($\theta$ = 0˚) (black line in Figure 1d )because there is reflection symmetry in the plane perpendicular to the propagation direction. When the incident angle is changed from $\theta$ = 45˚ to $\theta$ = -45˚, the CD spectra are reversed. $\theta$ is set as 45˚ in the following studies.

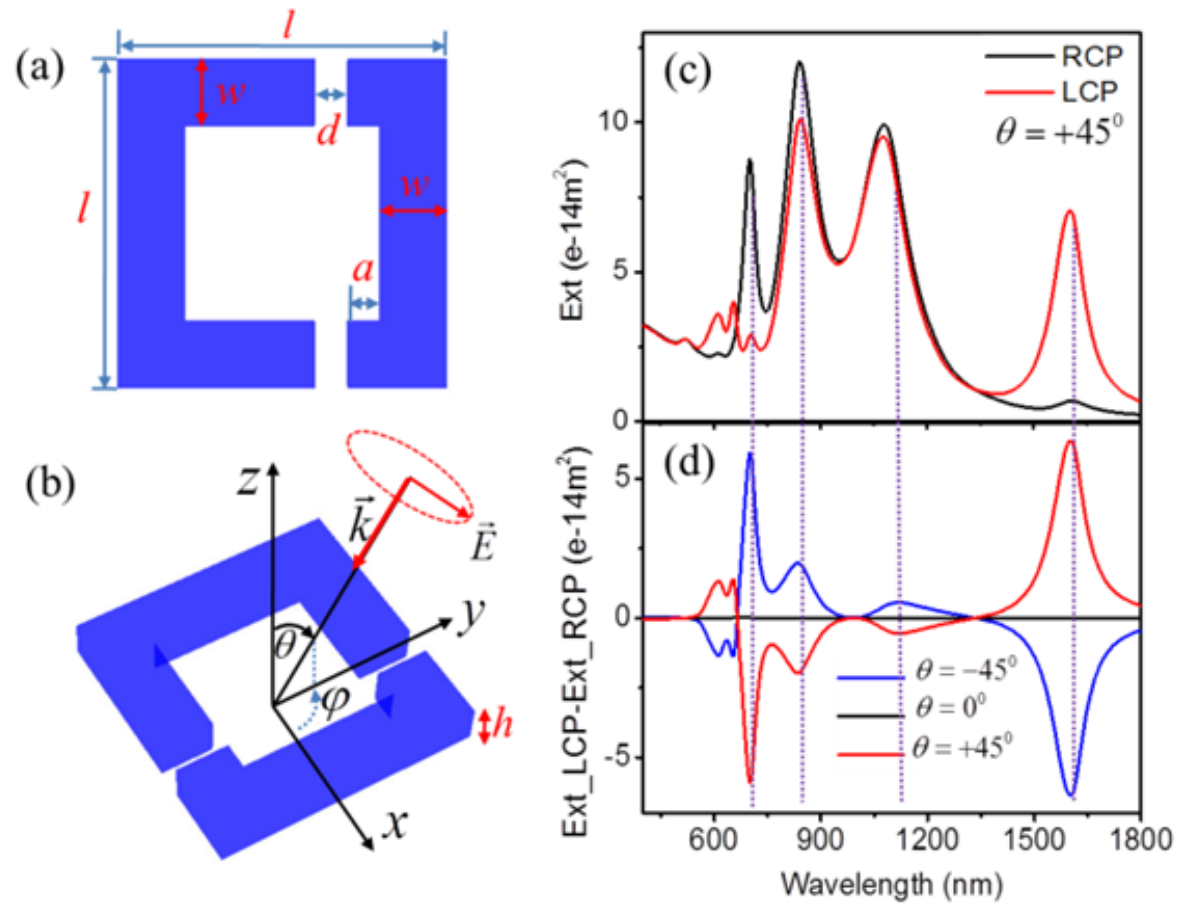

**Figure 1.** Fig (1) (a-b) Schematics of a splitting nanoring seen from different views with structure parameters labeled. The nanoring is located in the x-y plane and (b) gives the excitation case chosen in this paper, with y-axis along the connecting line of the two gaps. (c) Extinction spectra of the nanoring ($l$ = 200 nm, $w$ = 40nm, $w$ = 20nm, $h$ = 30nm, $a$ = 20nm) under LCP and RCP excited ($\theta$ = 45˚). (d) CD spectra of the nanoring illuminated with different incident angles ($\theta$ = 45˚(red), $\theta$ = 0˚ (black), $\theta$ = -45˚ (blue)).

**Hybridization of the modes and formation of the chiral response**

To easily draw out the analytical model later, we first analyze the surface plasmon modes supported by the splitting rectangle ring under different excitation polarizations (whose illumination direction is the same as in Figure 1b). Hybridization is a convenient way to understand the interaction of the structures.[30] Figure 2a shows the hybrid diagram of the rectangle ring under p polarized light ($\vec{E}$ in y-z plane). The left and right panels in Figure 2a are for the bigger and smaller parts individually, and the middle panel is for the whole ring structure. From the surface charge distributions one can easily find out that the four peaks of the whole ring are bonding modes and anti-bonding modes, in which peak 1, 3and 4 are bonding modes and peak 2 is anti-bonding modes. For the s polarized light ($\vec{E}$ along x), the condition is a bit different (Figure 2b). Because of the retardation in the light propagating direction, some modes oscillating perpendicular to x direction appear (peak 1', 2'', 3'') in the two individual parts. But from Figure 2b, one can still see that all peaks are bonding modes. From the surface charge distributions in Figure 2a and 2b, it is easy to find out that peak 1 and 3 are dark modes, where the current flows in a circle acting as a magnetron; peak 2 (see supporting information Figure S1)and 4 the bright modes and act as electric dipoles and quadruples.

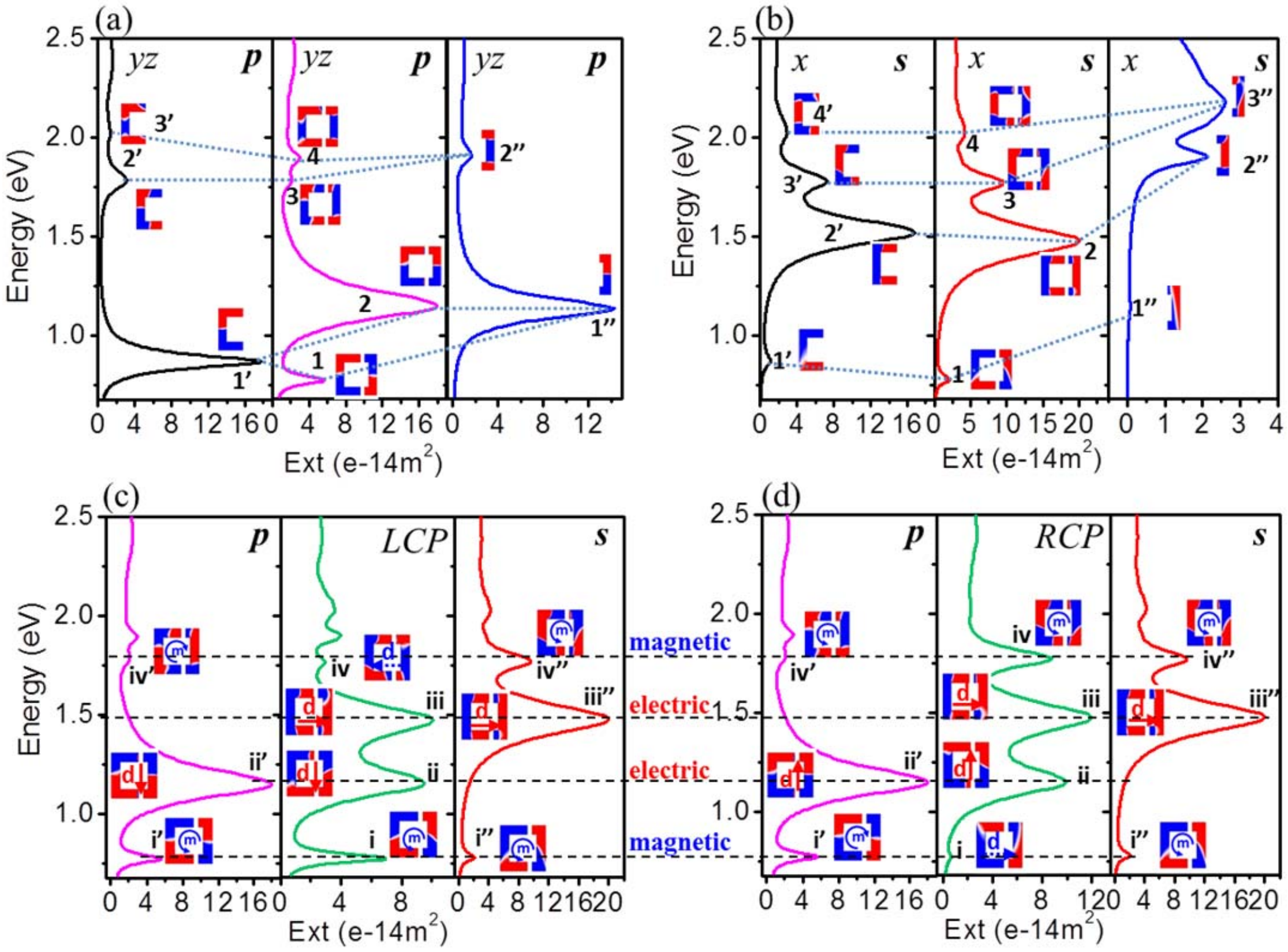

**Figure 2.** Hybridization diagrams (a, b) and mixing electric and magnetic modes under CPL(c, d) for the splitting rectangle ring with the same parameters as in Figure 1. Hybrid diagram of the structure under (a) p polarized light excitation and (b) s polarized light excitation. Modes analysis under LCP (c) and RCP (d) excitation with mixing p and s polarized light. Insets are surface charge distributions.

As the circularly polarized light can be decomposed into s and p polarized light with π/2 phase difference, we analyze the CPL situations by combining the results of s and p polarization excitations in Figure 2a and 2b, and in the following we will focus on the four main peaks (marked in Figure 2c and 2d). Figure 2c shows the condition for left-handed circularly polarized light (LCP), with the middle panel under LCP excitation, the left and right panels under linearly polarized light of p and s. From Figure 2c we can see that the four modes For LCP (i, ii, iii and iv) correspond to modes i’, ii’ and iv’ for p polarized light and i’’, iii’’ and iv’’ for s polarized light. The surface charge distributions for p polarized light is plotted in phase -π/2 while for s is 0, which is consistent with the incident light of LCP. For peak i, one can see that the surface charges for s and p light are in phase oscillating, so the superposition of the two modes will strengthen each other, which makes the peak i stronger. For peak ii and peak iii, as they are only from p or s light, so they are not weakened or strengthened from a *direct view* of this picture (the intensity change comes from the mixing of electric and magnetic mode which will be analyzed later). For peak iv, one can see from the surface charge distributions that iv’ and iv’’ are oscillating out of phase in opposite direction, so this peak is weakened in LCP; the effective oscillation is like a weak quadruple. Similarly, in Figure 2d, the condition for right-handed circularly polarized light (RCP) is shown. The surface charge distributions for p polarized light is plotted in phase π/2 while for s is 0, which is consistent with the incident light of RCP. Peak i is weakened and peak iv is strengthened, while peak ii and iii are somehow non-weakened and non-strengthened.

The diagrams in Figure 2c and 2d explain the difference for CD signals of the splitting rectangle ring structure, however, the origin of CD for peak ii and peak iii are not clear yet. For a better understanding of the origin of CD of the plasmonic structure, we propose an analytical model in the following and examine the theory in FEM model.

**Modes mixing for coupled electric and magnetic dipoles**

From Figure 2c and 2d, we know that the dark modes (peak i and peak iv) oscillate in a circular way, which is equivalent to a magneton and the bright modes (peak ii and peak iii) are electric dipoles. Analogic to chiral molecules, when the electric dipole and magnetic dipole can interact with each other in chiral plasmonic structures, where the electric dipole, magnetic dipole and wave vector construct a three-dimensional configuration that cannot overlap with its mirror form, no matter intrinsic or extrinsic chirality, there will be CD. For the symmetric configuration, even if there are strong magnetic modes, the electric dipole is always orthogonal to the magnetic dipole so that there is no CD. For a plasmonic meta-molecule with both magnetic and electric dipole modes excited, there is a mixed electric-magnetic dipole polarizability $G = G' + iG''$, which makes the electric dipole moment $p_e$ and magnetic dipole moment $p_m$ as[22]

$$\widetilde{\boldsymbol{p_e}} = \tilde{\alpha}\widetilde{\boldsymbol{E}} - i\tilde{G}\widetilde{\boldsymbol{B}},\ \widetilde{\boldsymbol{p_m}} = \tilde{\chi}\widetilde{\boldsymbol{B}} - i\tilde{G}\widetilde{\boldsymbol{E}}, (1)$$

where $\alpha = \alpha' + i\alpha''$ is the electric polarizability and $\chi = \chi' + i\chi''$ is the magnetic susceptibility. $\boldsymbol{E}$ and $\boldsymbol{B}$ are the local fields at the meta-molecule.

The extinction of the meta-molecule is

$$A^{\pm} = \frac{\omega}{2} Im\left(\widetilde{\boldsymbol{E}}^* \cdot \widetilde{\boldsymbol{p_e}} + \widetilde{\boldsymbol{B}}^* \cdot \widetilde{\boldsymbol{p_m}}\right) = \frac{\omega}{2}\left(\alpha''\left|\widetilde{\boldsymbol{E}}\right|^2 + \chi''\left|\widetilde{\boldsymbol{B}}\right|^2\right) + G''^{\pm}\omega\, Im(\widetilde{\boldsymbol{E}}^{\pm *} \cdot \widetilde{\boldsymbol{B}}^{\pm}), (2)$$

$$\Delta A = G''^{+}C^{+} - G''^{-}C^{-}, (3)$$

where $C = -\frac{\varepsilon_0\omega}{2} Im(\boldsymbol{E}^* \cdot \boldsymbol{B})$ is the electromagnetic field chirality[22]. $C_{CPL} = \pm\frac{\varepsilon_0\omega}{2c}E_0^2$ is the optical chirality for a right (−) or left (+) circularly polarized plane wave with electric field amplitude $E_0 = 1V/m$. Here we use $G''^{\pm}$ because the magnetic dipole is induced by plasmonic resonance, which is different for LCP and RCP, while for normal chiral molecules the intrinsic electric and magnetic polarizability is fixed. Following Schellman and Govorov [31, 32] we get that

$$G'' = -Im(\boldsymbol{p_e} \cdot (-i\boldsymbol{p_m})).\ (4)$$

Keeping the above in mend, we first investigate the system of coupled plasmonic electric and magnetic dipoles with coupled-dipole approximation method (see Supporting Information for details)[33, 34]. The dipole moments of the two coupled dipoles can be expressed as[35, 36]

$$\boldsymbol{p_e} = \varepsilon_0 \overleftrightarrow{\boldsymbol{\alpha_1}}\left(\boldsymbol{E_{1,in}} - Z_0 k^2 \overleftrightarrow{\boldsymbol{G_m}}(\boldsymbol{r_e}, \boldsymbol{r_m})\boldsymbol{p_m}\right), \quad (5)$$

$$\boldsymbol{p_m} = \overleftrightarrow{\boldsymbol{u_2}}\left(\boldsymbol{H_{2,in}} + ck^2 \overleftrightarrow{\boldsymbol{G_m}}(\boldsymbol{r_m}, \boldsymbol{r_e})\boldsymbol{p_e}\right), \quad (6)$$

Where $\overleftrightarrow{\boldsymbol{\alpha_1}}$ and $\overleftrightarrow{\boldsymbol{u_2}}$ are the polarizability tensors, and $\overleftrightarrow{\boldsymbol{G_m}}(\boldsymbol{r_j}, \boldsymbol{r_k})$ is the electric dyadic Green's function. Solving this set of two equations, we can obtain the self-consistent forms of dipole moments

$$\boldsymbol{p_e} = \frac{\varepsilon_0 \overleftrightarrow{\boldsymbol{\alpha_1}}\boldsymbol{E_{1,in}} - Z_0 k^2 \varepsilon_0 \overleftrightarrow{\boldsymbol{\alpha_1}}\overleftrightarrow{\boldsymbol{G_m}}(\boldsymbol{r_m}, \boldsymbol{r_e})\overleftrightarrow{\boldsymbol{u_2}}\boldsymbol{H_{2,in}}}{\overleftrightarrow{\boldsymbol{I}} + \boldsymbol{c}Z_0 k^4 \varepsilon_0 \overleftrightarrow{\boldsymbol{\alpha_1}}\overleftrightarrow{\boldsymbol{G_m}}(\boldsymbol{r_m}, \boldsymbol{r_e})\overleftrightarrow{\boldsymbol{u_2}}\overleftrightarrow{\boldsymbol{G_m}}(\boldsymbol{r_e}, \boldsymbol{r_m})}, \quad (7)$$

$$\boldsymbol{p_m} = \frac{\overleftrightarrow{\boldsymbol{u_2}}\boldsymbol{E_{2,in}} + \boldsymbol{c}k^2 \varepsilon_0 \overleftrightarrow{\boldsymbol{u_2}}\overleftrightarrow{\boldsymbol{G_m}}(\boldsymbol{r_e}, \boldsymbol{r_m})\overleftrightarrow{\boldsymbol{\alpha_1}}\boldsymbol{E_{1,in}}}{\overleftrightarrow{\boldsymbol{I}} + \boldsymbol{c}Z_0 k^4 \varepsilon_0 \overleftrightarrow{\boldsymbol{u_2}}\overleftrightarrow{\boldsymbol{G_m}}(\boldsymbol{r_e}, \boldsymbol{r_m})\overleftrightarrow{\boldsymbol{\alpha_1}}\overleftrightarrow{\boldsymbol{G_m}}(\boldsymbol{r_m}, \boldsymbol{r_e})}, \quad (8)$$

where $\overleftrightarrow{\boldsymbol{I}}$ is the unit dyad. Then the extinction of the system is

$$A^{\pm} = \frac{\omega}{2} Im(\boldsymbol{E}^* \cdot \boldsymbol{p_e} + \boldsymbol{B}^* \cdot \boldsymbol{p_m}), \quad (9)$$

Figure 3 shows the spectra of individual uncoupled dipoles and coupled dipoles under tilted LCP and RCP illuminations. The electric dipoles and magnetic dipoles are orthogonal in the incident plane as shown in Figure 3a inset. The polarizability of the electric dipoles $\overleftrightarrow{\boldsymbol{\alpha_1}}$ and the magnetic susceptibility of the magnetic $\overleftrightarrow{\boldsymbol{u_2}}$ are set based on an ellipsoid particle with long axis along x for electric dipole and long axis along z for magnetic dipole. The blue and red curves in Figure 3a are for individual electric and magnetic dipoles resonant at different wavelengths. When they are coupled together, one can see that the hybrid electric mode blue shifts and the magnetic mode red shifts as expected because of the hybridization. There is a dip right at the position of the individual magnetic resonant wavelength, which is known as Fano resonance. No surprisingly, the coupled system shows CD effect because of the mixed polarizability of the electric and magnetic dipoles just like the chiral molecules. To see a clearer picture, the dipole extinction power of the individual dipoles for the coupled system is plotted in Figure 3b. One can see that even if there is only electric (magnetic) moment for uncoupled dipoles, when they are coupled together, both the resonant peaks have electric and magnetic moments mixed. So there are interactions of electric and magnetic components in both electric and magnetic hybrid modes. Figure 3c shows CD spectra from a direct LCP and RCP extinction difference (red curve) and from formula 3 & 4 (blue curve). It can be seen that they matches very well with each other. This means that the generation mechanism of CD of the plasmonic chiral structures is quantitatively given and connected to the CD spectrum by the interaction and mixing of the electric and magnetic excitations.

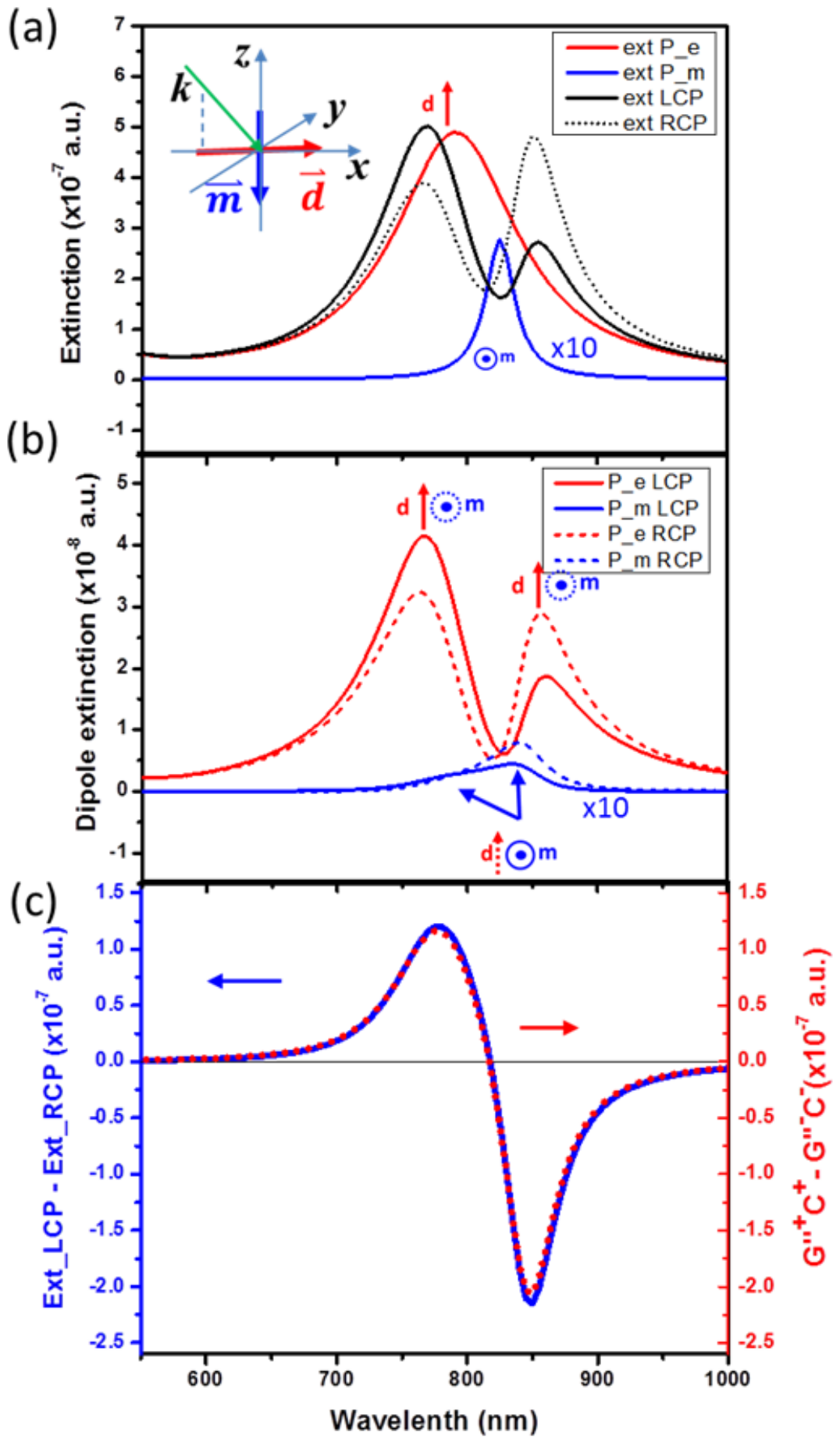

**Figure 3.** Coupled dipole approximation calculations for coupled electric dipole (ellipsoid a = 120 nm, b = c = 23 nm) and magnetic plasmonic dipole (ellipsoid a = b = 4.5 nm, c = 30 nm). (a) Extinction spectra for uncoupled electric and magnetic plasmonic dipoles, as well as coupled electric and magnetic plasmonic dipoles under CPL illuminations. (b) The dipole power of the individual coupled electric and magnetic plasmonic dipoles in (a). (c) Extinction difference (CD) of the coupled system (red curve) and imaginary part of the mixed electric and magnetic polarizability of the coupled system.

To verify the above mechanism, the plasmonic structure shown in Figure 1 is investigated with the same way as the analytical model but with electric and magnetic dipole obtained from numerical FEM results (Figure 4). Figure 4a shows the CD spectrum of an extinction difference for LCP and RCP.  The electric and magnetic dipole moments are obtained with[37] [classic electromagnetic dynamics, Jacksohn]

$$\boldsymbol{p_e} = \int d^3r'\boldsymbol{r}'\rho(\boldsymbol{r}'), \text{(10)}$$

$$\boldsymbol{p_m} = \frac{1}{2}\int d^3r'\ (\boldsymbol{r}' \times \boldsymbol{J}), \text{(11)}$$

where $\rho(\boldsymbol{r}')$ is the charge density and $\boldsymbol{J}(\boldsymbol{r}')$ is the current density. The derived electric and magnetic dipole power are plotted in Figure 4b. One can see that, very similar to the hybrid condition in Figure 3, the electric and magnetic dipole resonances are mixed together, which shows magnetic components at all resonant peaks. The magnetic dipole moment is much stronger in this hybrid structure because the split-ring resonators are able to tremendously enhance the total decay rate of the magnetic dipole emitters.[28] We should make it clear that the two peaks with the higher energy are multiple oscillations, which is out of the consideration of our model. We will discuss it later in the following. With the derived electric and magnetic dipole moments, CD spectrum can be calculated with Formula 3&4, shown in Figure 4c. It can be seen it shows itself in a very similar way to the curve in Figure 4a for the two lower energy modes. This confirms our analytical mechanism above in Figure 3 quantitatively, which is a further step for explanations of plasmonic chiral structures compared with the previous ones. However, it only works for dipole modes. For multiple modes, there are more than one equivalent magnetic dipoles oscillating out of phase, but with the delayed interaction with the electric modes, thus the whole effect is interference superposition[21]. While the hybrid model proposed here doesn't take the multiple and phase delay into consideration, so for the multiple modes, there is very big deviation (the left part of the black dot-dash line in Figure 4).

Because the electric dipole emitter and the induced magnetic dipole emitter have phase difference depending on wavelength, they will cause a Fano-type line shape as proved in other works.[21] As both the CD effect and Fano-type spectra are related to the electric and magnetic dipole (or dark mode) interactions, once there is CD effect in plasmonic structures, there is some Fano-type line shape in the spectrum.

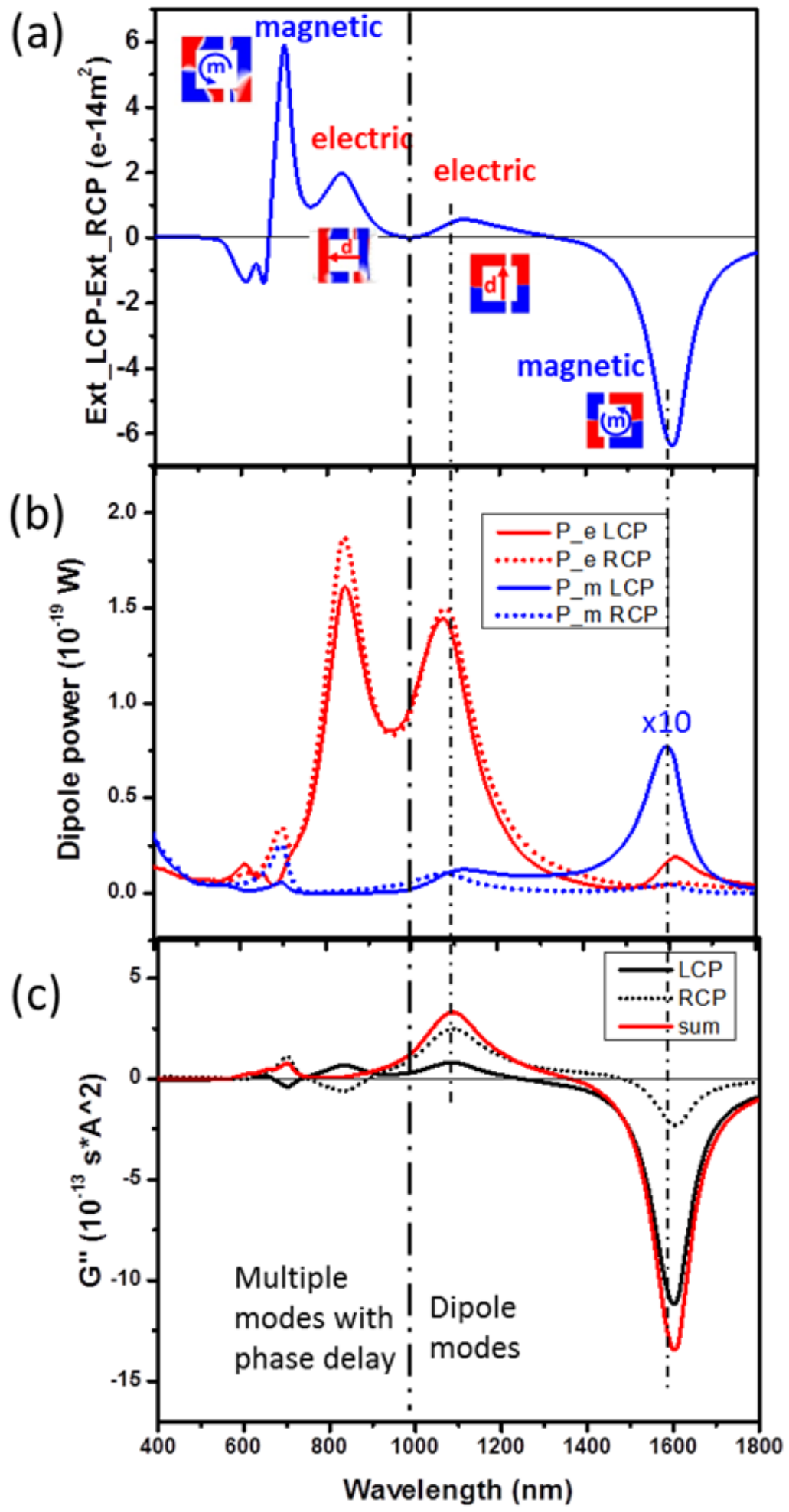

**Figure 4.** Coupled electric and magnetic dipoles analysis for the splitting rectangle ring shown in Figure 1. (a) Extinction difference (CD) of the structure. (b) The electric and magnetic dipole power yielded by the structure under CPL illumination. (c) Imaginary part of the mixed electric and magnetic polarizability of the structure under CPL illumination.

**Parameters dependent chirality**

To further investigate the CD of this structure, the dependence of the CD effect on several parameters of the system is studied as well, as shown in Figure 5. Figure 5a&b shows the extinction and CD spectra of the splitting rectangle rings excited by CPL light when the right asymmetric arm $a$ varies from 40 nm to 0 nm while the gap and the total length of the rectangle ring are kept constants (as shown in Figure 1a, $d$ = 20 nm, $l$ = 200 nm, $h$ = 30 nm, $w$=40 nm). When the right arm becomes shorter and the left one becomes longer, all hybrid modes red-shift except the ii modes (mainly domained by the shorter arm, as marked in figure 2) and the intensity increases ( Figure 5a&b). With the increasing of the left arm, the total magnetic mode becomes stronger, which will increase the CD of the structure. At the same time, the asymmetric Fano profile becomes more distinctive and thus induces the increasing of the CD response (Figure 5b).

The thickness of the system has a significant effect on the CD response. Splitting rectangle rings with thickness of 20 nm, 40 nm, and 60 nm are studied, as shown in Figure 5c&d, with other parameters remaining unchanged ($a$ = 20 nm, $d$ = 20 nm, $l$ = 200 nm, $w$=40 nm). From Figure 5c, it can be seen that all resonant peaks blue-shift and the extinction cross section of peak i increases under LCP excitation, but decreases under RCP excitation. When the thickness changes to 60 nm, the structure exhibits totally left handedness. This is because when the cross section of the bar increases, the induced current and dipole moment will increase, so the corresponding CD response is enhanced markably with the thickness increasing and then with the stronger interplay of the electric and magnetic dipoles (Figure 5d). This stronger and stronger interaction causes a total CPL selection in the lowest mode, which is seldom observed in plasmonic chirality.

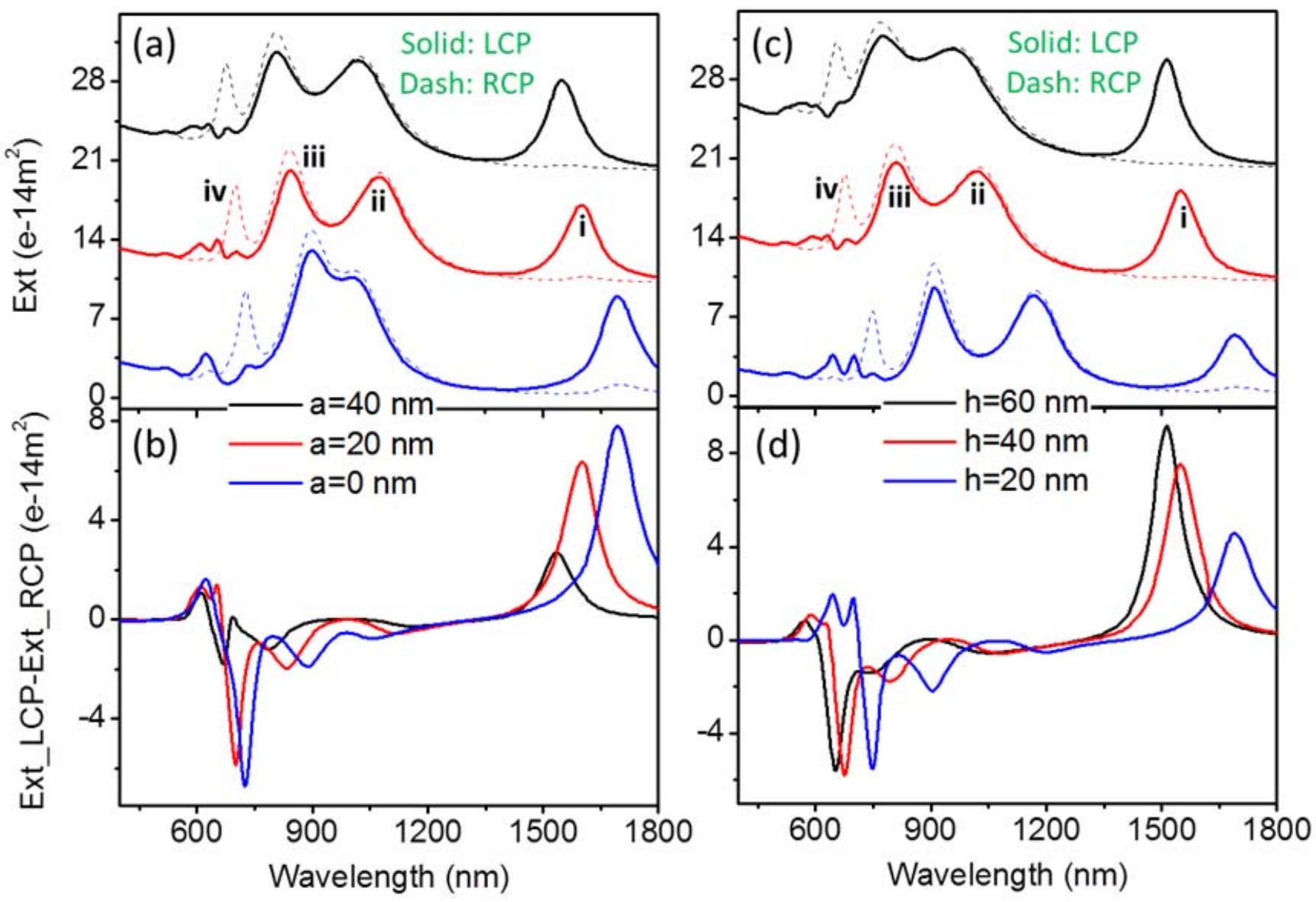

**Figure 5.** Dependence of the CD effect on parameters a and h (see definitions in Figure 1). Top: Extinction spectra of (a) splitting rectangle ring with different asymmetric arms and (c) Splitting rectangle ring with different thicknesses. To make a better perspective, every spectrum in this part has an offset of $10*10^{-14}$ m$^2$ than the one below it. (b, d) CD spectra of the above splitting rectangle rings. Solid lines in upper row are for LCP and dashed lines are for RCP.

### Chiral near field properties and sensing

One of the most important issues of plasmonic chiral structures is the chiral near field enhancement for chiral molecule detection and sensing. For chiral molecules, the CD signal follows the same relation as formula 3, but with fixed mixed electric and magnetic polarizability G''[5]

$$\Delta A = G''(C^+ - C^-) = 2G''|C_{CPL}|$$

For plasmonic structures enhancing molecule chiral detection,[25] the integration of the local chiral field where the molecules adsorb determines the CD signal. The enhancement factor for chirality of the field $\hat{C} = C/|C_{CPL}|$ is considered in the following. The whole chiral field enhancement is the volume averaged chiral spectra where the molecule is located and is calculated with [26]$\langle \hat{C} \rangle = \frac{1}{V}\int_V \hat{C} \cdot dV$.

We first investigate the chiral near field distributions of the same structure as in Figure 1 (Figure 6). Figure 6a shows the electric field enhancement at the four resonant peaks under RCP and LCP excitation. It is not surprising that the fields in the splitting gaps are obviously enhanced. An interesting thing is that the field enhancment in the two gaps has different selectivity to the handedness of the exciting light and to the resonant modes. For instance, the electric field in the upper gap is stronger at 1610 nm, 700 nm for RCP and 1080 nm for LCP; the electric field in the lower gap is stronger at 1080 nm for RCP. This may have potential applications in selective photo-catalysis. In addition, a more interesting and important thing is that the chiral field in the gaps are also obviously selective for LCP and RCP as shown in Figure 6b. One can see that the upper gap enhances the RCP chiral field at 1610 nm, 840 nm and LCP chiral field at 1080 nm, 840 nm. The lower gap enhances the RCP chiral field at 1080 nm and LCP chiral field at 1610 nm. If one focuses on one resonant peak, e.g. 1610 nm peak, RCP light will excite strong chiral field in the upper gap while LCP light excites stronger chiral field in the lower gap. The case is opposite for 1080 nm peak. The selective switching enhancement of the chiral field for CPL is very useful in chiral molecule sensing and catalysis. In most traditional plasmonic structures for chiral field enhancement, the target gaps have enhancement effect for both LCP and RCP, while here, in the resonant peak 1610 nm the upper gap keeps $C^-$ field and the lower gap keeps $C^+$ field; in the resonant peak 1080 nm, the upper gap keeps $C^+$ field and the lower gap keeps $C^-$ field.

In real applications, one usually hopes the left-handed chiral molecules enhanced by $C^+$ field and the right-handed chiral molecules enhanced by $C^-$ field. Here this selective enhancement for chiral fields may be useful for catalyzing chiral chemical reactions or detection simultaneously.

For a direct and clear picture to see this selective chiral fields, volume averaged chiral field enhancement $\langle \hat{C} \rangle$ spectra in the upper gap (Vol 1) and lower gap (Vol 2) are plotted in Figure 7. From the spectra one can clearly see that for the mode 1610 nm, the chiral field in Vol 1 is always negative, and in Vol 2 is always positive; while for the mode 1080 nm, the case is reversed. The larger difference for mode 1610 nm is more important and applicable in practice.

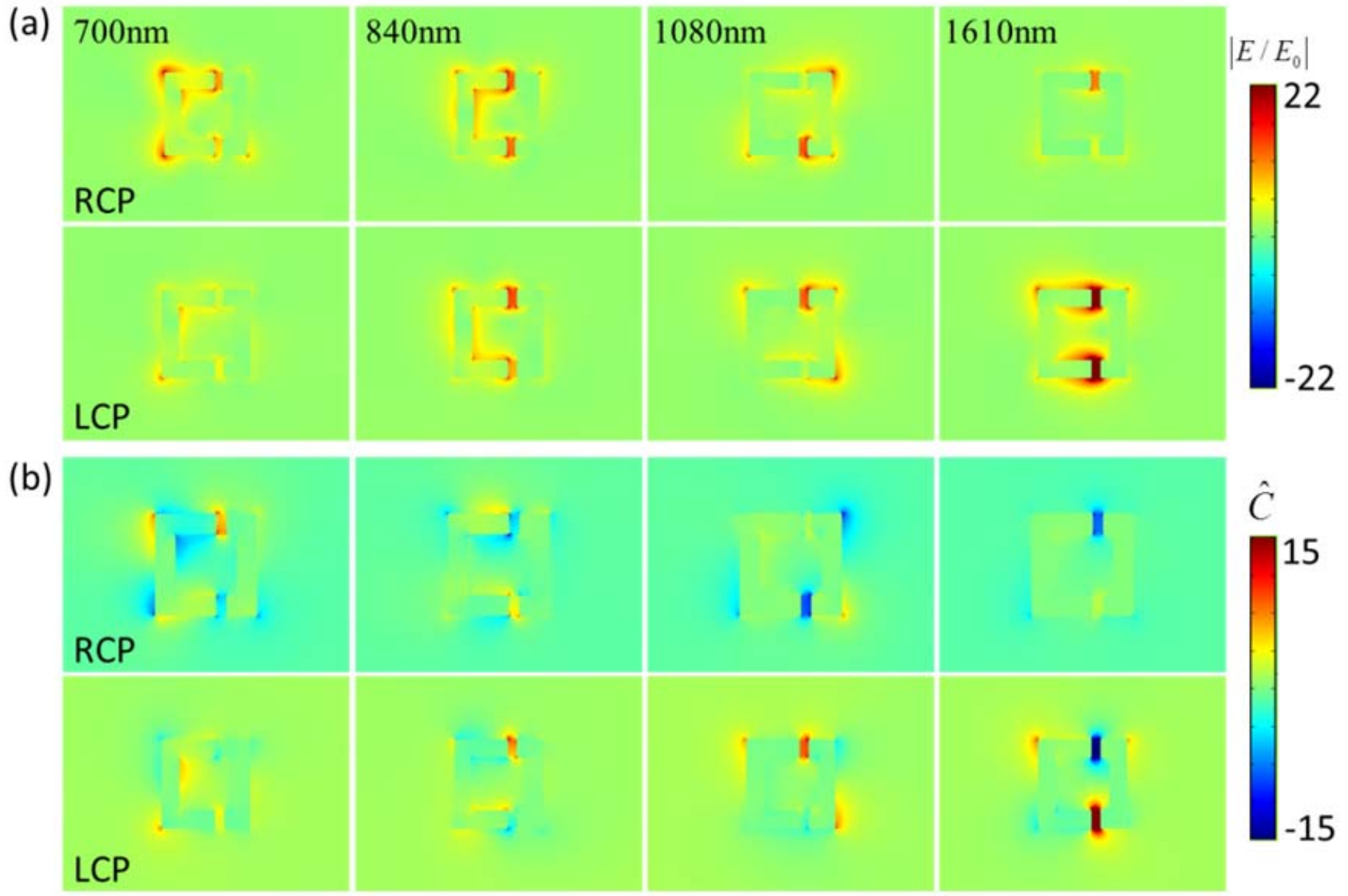

**Figure 6.** Enhancement distributions of field intensity and chirality. (a) Electric field enhancement and (b) Chiral field enhancement of the structure in Figure 1 at the four resonant peaks under RCP and LCP illumination. The slices are cut in the middle of the thickness.

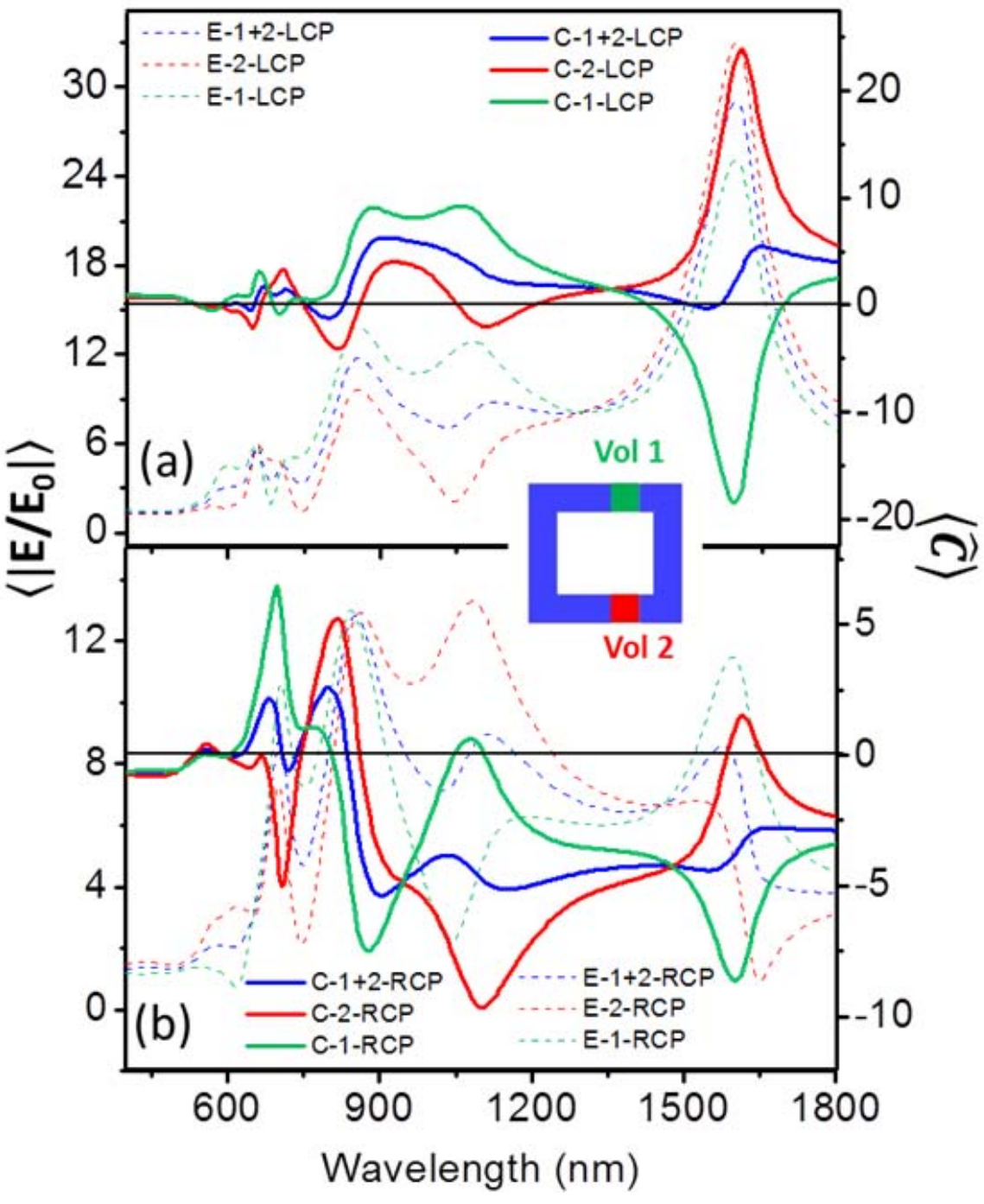

**Figure 7.** Volume averaged chiral field enhancement under LCP (a) and RCP (b) illumination.

**Conclusion and discussion**

In this work, the generation mechanism of extrinsic plasmonic chirality is quantitatively presented by an analytical model of coupled plasmonic electric and magnetic dipoles. The strong interplay of electric and induced magnetic dipoles will cause a mixed electric and magnetic polarizability, which is responsible for the CD of the meta-molecule to LCP and RCP light. The model is verified with numerical FEM results in splitting rectangle rings. The results show that the circular current yielded in the splitting rectangle rings behaves like a magneton. The hybridization of the electric mode and magnetic mode results in a mixing of the two modes, coincident with the above analytical model. Parameters dependent CD response of splitting rectangle rings is also investigated. The analysis of the chiral near field of the structure shows potential applications in chiral molecule sensing.

According to the molecular optical activity theory, the quantification of the mechanism of the plasmonic CD effect by both analytical and numerical model is expected to be applicable to all of the plasmonic extrinsic chiral structures, which is a continuous issue in the future. Expanding the model to higher order modes is necessary and needs more work, since higher order modes may exhibits stronger CD [21].

**Methods**

**FEM simulation:** All full wave numerical simulation were done by using finite element method (FEM, commercial software package, Comsol Multiphysics 4.3a). The Au splitting rectangle ring were put in a homogeneous surrounding medium of effective refractive index 1.1. Non-uniform meshes were used for formatting the object. The largest mesh was set less than $\lambda/6$. Perfect matched layer (PML) was used to minimize the scattering from the outer boundary. The nanoring was put in x-y plane. The incident light was set to 1V/m and propagates in the y-z plane with off the z axis. The total scattering cross sections were obtained by integrating the scattered power flux over an enclosed surface outside the nanoring, while the absorption cross sections were determined by integrating the Ohmic heating within the nanoring. The circular dichroism of the system were calculated as the difference in extinction under left and right handed circularly polarized light ($CD = \sigma_L - \sigma_R$). The super chiral field was plotted with $\hat{C} = C/|C_{CPL}|$, where $C$ is defined as $C = -\frac{\varepsilon_0 \omega}{2} Im(\boldsymbol{E}^* \cdot \boldsymbol{B})$, and $C_{CPL} = \pm \frac{\varepsilon_0 \omega}{2c} E_0^2$. The volume averaged chiral spectra were got with $\langle \hat{C} \rangle = \frac{1}{V} \int_V \hat{C} \cdot dV$

**The coupled dipole approximation method:** It is too long to put here, please see supporting information for details.

**Acknowledgement**

This work was supported by the National Natural Science Foundation of China (11204390), Natural Science Foundation Project of CQ CSTC (2014jcyjA40002), Science and Technology Project of CQ

Commission of Education (KJ1500636) and Special Fund for Agro-scientific Research in the Public Interest (201303045).

**Author contributions**

Y. Fang launched and supervised the project. L. Hu did the FEM simulations, X. Tian wrote the expressions of electric and magnetic dipoles for FEM. Y. Fang proposed the analytical dipole model and wrote the Matlab codes. Y. Fang and X. Tian discussed the coupling mechanism of dipoles. Y. Fang analyzed the all of the data and wrote the paper. All of the authors contributed useful suggestions and revised the paper.

**Competing financial interests:** The authors declare no competing financial interests.

**Supporting Information** Available for near field distributions of magnetic field of the splitting rectangle ring, coupled electric and magnetic dipoles method.

Supplementary Information for

## Quantitatively analyzing the mechanism of giant circular dichroism in extrinsic plasmonic chiral nanostructures by the interplay of electric and magnetic dipoles

Li Hu[1,4,†], Xiaorui Tian[2,†], Yingzhou Huang[1], Xinqiang Wang[1] and Yurui Fang[3,*]

[1]Soft Matter and Interdisciplinary Research Center, College of Physics, Chongqing University, Chongqing, 400044, P. R. China

[2]College of Chemistry, Chemical Engineering and Materials Science, Shandong Normal University, Jinan 250014, China

[3]Bionanophotonics, Department of Applied Physics, Chalmers University of Technology, Göteborg, SE-41296, Sweden.

[4]School of Computer Science and Information Engineering, Chongqing Technology and Business University, Chongqing, 400067, China

†These authors contribute equally

*Corresponding Email: yurui.fang@chalmers.se (Y. Fang)

### The electric and magnetic dyadic Green's functions

The electric dyadic Green's tensor $\overleftrightarrow{\boldsymbol{G_e}}(\boldsymbol{r},\boldsymbol{r_0})$ and magnetic dyadic Green's function tensor $\overleftrightarrow{\boldsymbol{G_m}}(\boldsymbol{r},\boldsymbol{r_0})$ are the free space field susceptibility tensors propagator relating an electric dipole source $\boldsymbol{p_e}$ at position $\boldsymbol{r_0}$ in vacuum to the electric field $\boldsymbol{E}$ and magnetic field **H** it generates at position $\boldsymbol{r}$ through

$$\boldsymbol{E}(\boldsymbol{r}) = \frac{k^2}{\varepsilon_0}\overleftrightarrow{\boldsymbol{G_e}}(\boldsymbol{r},\boldsymbol{r_0})\boldsymbol{p_e} \quad (1)$$

$$\boldsymbol{H}(\boldsymbol{r}) = ck^2\overleftrightarrow{\boldsymbol{G_m}}(\boldsymbol{r},\boldsymbol{r_0})\boldsymbol{p_e} \quad (2)$$

For the electric and magnetic fields generated by a magnetic dipole $\boldsymbol{p_m}$,

$$\boldsymbol{E}(\boldsymbol{r}) = -Z_0k^2\overleftrightarrow{\boldsymbol{G_m}}(\boldsymbol{r},\boldsymbol{r_0})\boldsymbol{p_m} \quad (3)$$

$$\boldsymbol{H}(\boldsymbol{r}) = k^2\overleftrightarrow{\boldsymbol{G_e}}(\boldsymbol{r},\boldsymbol{r_0})\boldsymbol{p_m} \quad (4)$$

With

$$\overleftrightarrow{\boldsymbol{G_e}}(\boldsymbol{r},\boldsymbol{r_0}) = \frac{e^{ikr}}{r}\left[\left(\hat{\boldsymbol{n}}\otimes\hat{\boldsymbol{n}}-\overleftrightarrow{\boldsymbol{I}}\right)+\frac{ikr-1}{k^2r^2}\left(3\cdot\hat{\boldsymbol{n}}\otimes\hat{\boldsymbol{n}}-\overleftrightarrow{\boldsymbol{I}}\right)\right] \quad (5)$$

$$\overleftrightarrow{G_m}(\boldsymbol{r},\boldsymbol{r_0}) = \frac{e^{ikr}}{r}\left(1+\frac{i}{kr}\right)(\widehat{\boldsymbol{n}}\times\overleftrightarrow{\boldsymbol{I}}) \quad (6)$$

$$\overleftrightarrow{G_e}(\boldsymbol{r},\boldsymbol{r_0})\boldsymbol{p} = \frac{1}{4\pi}\frac{e^{ikr}}{r}\left[(\widehat{\boldsymbol{n}}\times\boldsymbol{p})\times\widehat{\boldsymbol{n}}+\frac{ikr-1}{k^2r^2}(3\cdot\widehat{\boldsymbol{n}}(\widehat{\boldsymbol{n}}\cdot\boldsymbol{p})-\boldsymbol{p})\right] \quad (7)$$

$$\overleftrightarrow{G_m}(\boldsymbol{r},\boldsymbol{r_0})\boldsymbol{p} = \frac{e^{ikr}}{r}\left(1+\frac{i}{kr}\right)(\widehat{\boldsymbol{n}}\times\boldsymbol{p}) \quad (8)$$

Where $r = |\boldsymbol{r}-\boldsymbol{r_0}|$, $k = 2\pi/\lambda$ and $\widehat{\boldsymbol{n}} = \frac{r-r_0}{r}$.

**The coupled dipole approximation method**

Let us consider many three dimensional dipole scatters.

The local field at each dipole can be expressed as[1]

$$\boldsymbol{p_{e,j}} = \varepsilon_0\overleftrightarrow{\boldsymbol{\alpha_j}}\boldsymbol{E_{j,total}} = \varepsilon_0\overleftrightarrow{\boldsymbol{\alpha_j}}\left(\boldsymbol{E_{j,in}}+\sum_{k=1,k\neq j}^{N}\left(\frac{k^2}{\varepsilon_0}\overleftrightarrow{G_e}(\boldsymbol{r_j},\boldsymbol{r_k})\boldsymbol{p_{e,k}}-Z_0k^2\overleftrightarrow{G_m}(\boldsymbol{r},\boldsymbol{r_0})\boldsymbol{p_{m,k}}\right)\right) \quad (9)$$

$$\boldsymbol{p_{m,j}} = \overleftrightarrow{\boldsymbol{u_j}}\boldsymbol{H_{j,total}} = \overleftrightarrow{\boldsymbol{u_j}}\left(\boldsymbol{H_{j,in}}+\sum_{k=1,k\neq j}^{N}\left(ck^2\overleftrightarrow{G_m}(\boldsymbol{r_j},\boldsymbol{r_k})\boldsymbol{p_{e,k}}+k^2\overleftrightarrow{G_e}(\boldsymbol{r},\boldsymbol{r_0})\boldsymbol{p_{m,k}}\right)\right) \quad (10)$$

For coupled electric and magnetic dipoles, we have

$$\boldsymbol{p_e} = \varepsilon_0\overleftrightarrow{\boldsymbol{\alpha_1}}\left(\boldsymbol{E_{1,in}}-Z_0k^2\overleftrightarrow{G_m}(\boldsymbol{r_e},\boldsymbol{r_m})\boldsymbol{p_m}\right) \quad (11)$$

$$\boldsymbol{p_m} = \overleftrightarrow{\boldsymbol{u_2}}\left(\boldsymbol{H_{2,in}}+ck^2\overleftrightarrow{G_m}(\boldsymbol{r_m},\boldsymbol{r_e})\boldsymbol{p_e}\right) \quad (12)$$

We can easily get the self-consistent form of dipole moments

$$\boldsymbol{p_e} = \frac{\varepsilon_0\overleftrightarrow{\boldsymbol{\alpha_1}}\boldsymbol{E_{1,in}}-Z_0k^2\varepsilon_0\overleftrightarrow{\boldsymbol{\alpha_1}}\overleftrightarrow{G_m}(\boldsymbol{r_m},\boldsymbol{r_e})\overleftrightarrow{\boldsymbol{u_2}}\boldsymbol{H_{2,in}}}{\overleftrightarrow{\boldsymbol{I}}+cZ_0k^4\varepsilon_0\overleftrightarrow{\boldsymbol{\alpha_1}}\overleftrightarrow{G_m}(\boldsymbol{r_m},\boldsymbol{r_e})\overleftrightarrow{\boldsymbol{u_2}}\overleftrightarrow{G_m}(\boldsymbol{r_e},\boldsymbol{r_m})} \quad (13)$$

$$\boldsymbol{p_m} = \frac{\overleftrightarrow{\boldsymbol{u_2}}\boldsymbol{E_{2,in}}+ck^2\varepsilon_0\overleftrightarrow{\boldsymbol{u_2}}\overleftrightarrow{G_m}(\boldsymbol{r_e},\boldsymbol{r_m})\overleftrightarrow{\boldsymbol{\alpha_1}}\boldsymbol{E_{1,in}}}{\overleftrightarrow{\boldsymbol{I}}+cZ_0k^4\varepsilon_0\overleftrightarrow{\boldsymbol{u_2}}\overleftrightarrow{G_m}(\boldsymbol{r_e},\boldsymbol{r_m})\overleftrightarrow{\boldsymbol{\alpha_1}}\overleftrightarrow{G_m}(\boldsymbol{r_m},\boldsymbol{r_e})} \quad (14)$$

The extinction is

$$A = \frac{\omega}{2} Im(\boldsymbol{E}^* \cdot \boldsymbol{p_e} + \boldsymbol{B}^* \cdot \boldsymbol{p_m}) \quad (15)$$

**Radiation power of the dipoles**

The radiation power expressions of the electric dipole $\boldsymbol{p_e}$ and magnetic dipole $\boldsymbol{p_m}$ are

$$Q_e = \frac{\omega^4}{12\pi\varepsilon_0 c^3} |\boldsymbol{p_e}|^2 \quad (16)$$

$$Q_m = \frac{\omega^4 Z_0}{12\pi c^4} |\boldsymbol{p_m}|^2 \quad (17)$$

**The polarizability of the ellipsoid dipole**

For an ellipsoid the polarizability tensor is

$$\overleftrightarrow{\boldsymbol{\alpha}}(\boldsymbol{r},\omega) = \overleftrightarrow{\boldsymbol{\alpha_0}}(\boldsymbol{r},\omega)[\overleftrightarrow{\boldsymbol{I}} - \left(\frac{2}{3}\right) i k_0^3 \overleftrightarrow{\boldsymbol{\alpha_0}}(\boldsymbol{r},\omega) - k^2/\overleftrightarrow{\boldsymbol{\alpha_0}}]^{-1} \quad (18)$$

where $\overleftrightarrow{\boldsymbol{\alpha_0}}(\boldsymbol{r},\omega)$ is the Clausius-Mossotti polarizability

$$\overleftrightarrow{\boldsymbol{\alpha_0}} = \begin{pmatrix} \alpha_1 & 0 & 0 \\ 0 & \alpha_2 & 0 \\ 0 & 0 & \alpha_3 \end{pmatrix} \quad (19)$$

$$\alpha_j = 4\pi abc \frac{(\boldsymbol{\varepsilon_{particle}} - \boldsymbol{\varepsilon_{medium}})}{(3\boldsymbol{\varepsilon_{particle}} + 3L_j(\boldsymbol{\varepsilon_{particle}} - \boldsymbol{\varepsilon_{medium}}))} \quad (20)$$

$$L_x = \frac{abc}{2}\int_0^\infty \frac{dq}{(x^2+q)f(q)}$$

with $x = a, b, c$, $f(q) = [(q+a^2)(q+b^2)(q+c^2)]^{1/2}$ and $a, b, c$ are the axis of the ellipsoid[2].

$\overleftrightarrow{\boldsymbol{u}}$ is obtained in the same way. Because because nature material has bad magnetic response in optical frequency and the magnetons in our paper are yield by the plasmon resonance with circular current, we used a fake $\boldsymbol{u_{particle}}$ value, which is the same as $\boldsymbol{\varepsilon_{Au}}$ but with imaginary part divided by 1.5 (the magnetic mode is dark, so the spectrum profile is narrower).

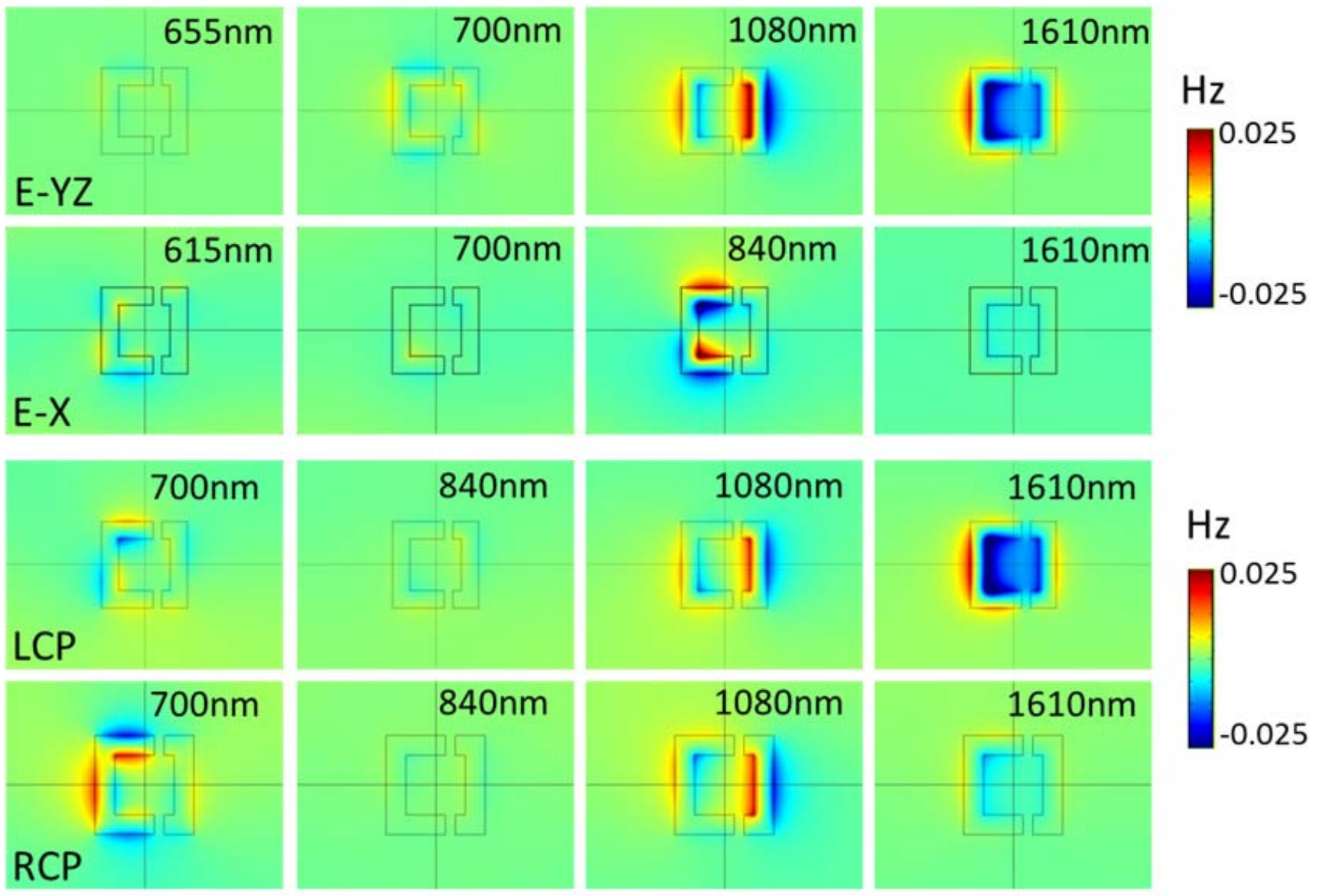

**Figure S1**. The magnetic field distributions in z direction for the simulations in Figure 2.